\documentclass[prd,aps,twocolumn,a4paper,showkeys,nofootinbib]{revtex4-1}

\usepackage{graphicx,psfrag}
\usepackage{mathrsfs}
\usepackage{amsmath,amsfonts,amssymb}
\usepackage{multirow}
\usepackage{comment}

\usepackage[normalem]{ulem}
\usepackage{hyperref}
\usepackage{enumitem}
\usepackage{acronym}
\usepackage{lmodern}
\usepackage[utf8]{inputenc}
\usepackage[T1]{fontenc}
\usepackage{url}

\newcommand{\be}{\begin{equation}}
\newcommand{\ee}{\end{equation}}
\newcommand{\bea}{\begin{eqnarray}}
\newcommand{\eea}{\end{eqnarray}}
\newcommand{\bel}{\begin{align}}
\newcommand{\eel}{\end{align}}

\def\lm{{\ell m}}

\def\e{{\rm e}}
\def\i{{\rm i}}
\def\d{{\rm d}}

\def\Msun{{\rm M_{\odot}}}
\def\GMc2{{\rm G M_{\odot} c^{-2}}}
\def\Mpc{{\rm Mpc}}

\def\M{\mathcal{M}}

\def\Mo{{\rm M_{\odot}}}
\def\Mc{\mathcal{M}}
\def\Mecc{\Mc_{\rm ecc}}
\def\chieff{\chi_{\rm eff}}
\def\chip{\chi_{\rm p}}
\def\kt2{\kappa^\text{T}_2}

\def\tLam{\tilde{\Lambda}}
\def\params{\boldsymbol{\theta}}

\def\ie{\textit{i.e.}}
\def\eg{\textit{e.g.}}

\usepackage{pifont} 

\newcommand{\bajes}{\texttt{bajes}}
\newcommand{\TEOB}[1]{\texttt{TEOBResumS}\texttt{#1}}
\newcommand{\dali}{{Dal\'i}}

\usepackage{color}
\definecolor{cyan}{rgb}{0,0.9,0.9}
\definecolor{orange}{rgb}{0.9,0.5,0}
\definecolor{magenta}{rgb}{1,0,1}
\definecolor{purple}{rgb}{0.8,0.4,0.8}
\definecolor{gray}{rgb}{0.8242,0.8242,0.8242}

\newacro{bbh}[BBH]{binary black hole}
\newacro{bh}[BH]{black hole}
\newacroplural{bh}[BHs]{black holes}
\newacro{bhns}[BHNS]{black hole-neutron star}
\newacro{bns}[BNS]{binary neutron star}
\newacro{bf}[BF]{Bayes' factor}
\newacro{cbc}[CBC]{compact binary coalescence}
\newacro{ci}[CI]{credible interval}
\newacro{ce}[CE]{Cosmic Explorer}
\newacro{da}[DA]{data analysis}
\newacro{et}[ET]{Einstein Telescope}
\newacro{emri}[EMRI]{extreme mass ratio inspiral}
\newacro{eob}[EOB]{effective-one-body}
\newacro{eos}[EoS]{equation of state}
\newacro{eom}[EOM]{equations of motion}
\newacro{fd}[FD]{frequency domain}
\newacro{fft}[FFT]{Fast Fourier transform}
\newacro{gw}[GW]{gravitational wave}
\newacro{gr}[GR]{general relativity}
\newacro{grb}[GRB]{gamma-ray burst}
\newacro{grhd}[GRHD]{general-relativistic hydrodynamics}
\newacro{gwosc}[GWOSC]{Gravitational Wave Open Science Center}
\newacro{gwtc1}[GWTC-1]{the first gravitational-wave transients catalog}
\newacro{gsf}[GSF]{Gravitational Self Force}
\newacro{hm}[HM]{Higher mode}
\newacroplural{hm}[HMs]{Higher modes}
\newacro{ifo}[IFO]{interferometer}
\newacro{imr}[IMR]{inspiral-merger-ringdown}
\newacro{im}[IM]{inspiral-to-merger}
\newacro{kagra}[KAGRA]{Kamioka Gravitational Wave Detector}
\newacro{ligo}[LIGO]{Laser Interferometer Gravitational-Wave Observatory}
\newacro{lso}[LSO]{Last Stable Orbit}
\newacro{lvc}[LVC]{LIGO-Virgo Collaboration}
\newacro{lvk}[LVK]{LIGO-Virgo-Kagra Collaboration}
\newacro{lo}[LO]{leading order}
\newacro{ns}[NS]{neutron star}
\newacroplural{ns}[NSs]{neutron stars}
\newacro{nr}[NR]{numerical relativity}
\newacro{nqc}[NQCs]{next-to-quasicircular corrections}
\newacro{nlo}[NLO]{next-to-leading order}
\newacro{nnlo}[NNLO]{next-to-next-to-leading order}
\newacro{n3lo}[N3LO]{next-to-next-to-next-to-leading order}
\newacro{n4lo}[N3LO]{next-to-next-to-next-to-next-to-leading order}
\newacro{ode}[ODE]{Ordinary Differential Equation}
\newacroplural{ode}[ODEs]{Ordinary Differential Equations}
\newacro{pe}[PE]{parameter estimation}
\newacro{pn}[PN]{post-Newtonian}
\newacro{pm}[PM]{post-merger}
\newacro{psd}[PSD]{power spectral density}
\newacro{pa}[PA]{post-adiabatic}
\newacro{qnm}[QNM]{quasi-normal mode}
\newacro{qc}[QC]{quasicircular}
\newacro{sm}[SM]{Supplemental Material}
\newacro{snr}[SNR]{signal-to-noise ratio}
\newacro{spa}[SPA]{stationary-phase approximation}
\newacro{sxs}[SXS]{Simulating eXtreme Spacetimes}
\newacro{td}[TD]{time domain}
\newacro{xg}[XG]{Next Generation}

\begin{document}

\title{%
  Gravitational waves from eccentric binary neutron star mergers:\\
  Systematic biases and inadequacy of quasicircular templates
}

\author{Giulia \surname{Huez}$^{1}$}
\author{Sebastiano \surname{Bernuzzi}$^{1}$}
\author{Matteo \surname{Breschi}$^{1}$}
\author{Rossella \surname{Gamba}$^{2,3}$}

\affiliation{${}^1$Theoretisch-Physikalisches Institut, Friedrich-Schiller-Universit{\"a}t Jena, 07743, Jena, Germany \\
            ${}^2$Institute for Gravitation $\&$ the Cosmos, The Pennsylvania State University, University Park PA 16802, USA \\
            ${}^3$Department of Physics, University of California, Berkeley, CA 94720, USA}

\date{\today}

\begin{abstract}
  The use of quasicircular waveforms in matched-filter analyses of signals from 
  eccentric \ac{bns} mergers can lead to biases in the
  source's parameter estimation.
  We demonstrate that significant biases can be present already for moderate eccentricities $e_{0}\gtrsim0.05$ and signals detected by LIGO-Virgo-KAGRA with \ac{snr} ${\gtrsim}12$.
  We perform systematic Bayesian mock analyses of unequal-mass nonspinning \ac{bns} signals up to
  eccentricities $e_0\sim0.1$ using quasicircular \ac{eob} waveforms with spins.
  We find fractional \ac{snr} losses up to tens of percent and up to 16$\sigma$ deviations in the inference of the chirp mass. The latter effect is sufficiently large to lead to an incorrect (and ambiguous) source identification. 
  The inclusion of spin precession in the quasicircular waveform does not capture eccentricity effects.
  We conclude that high-precision observations with advanced (and next generation) detectors are likely to require standardized, accurate, and fast eccentric waveforms.
\end{abstract}


\maketitle

\section{Introduction} 

The observed \ac{bns} in the Milky Way have a wide
range of eccentricities~\cite{Zhu:2017znf,Andrews:2019vou}.  
A short-period, highly eccentric
sub-population that includes the Hulse-Taylor binary has eccentricity
up to ${\sim}0.8$. By the time of merger, field binaries are expected
to circularize and to enter the sensitivity band of ground-based
\ac{gw} detectors at about 10~Hz with negligible eccentricity
$e_0\lesssim 10^{-4}$~\cite{Peters:1964zz,Kowalska:2010qg}.  
However, in case a highly eccentric
subpopulation is formed in a more dynamical environment, the binaries
might retain a non-negligible eccentricity at merger~\cite{Lower:2018seu}. 
\ac{bns} searches with quasicircular templates are expected to be effectual 
up to $e_0\lesssim0.05$~\cite{Huerta:2013qb,DallAmico:2025eov}.
A recent search for eccentric signals placed a 90\% upper limit of
${\sim}1700$ mergers ${\rm Gpc}^{-3}{\rm Yr}^{-1}$ for eccentricities
${\lesssim} 0.43$ at 10~Hz~\cite{Nitz:2019spj,Dhurkunde:2023qoe} (cf.
\cite{Phukon:2024amh,Kacanja:2024pjh,Lenon:2021zac,Pal:2023dyg}
for further eccentric template banks and search algorithms and
\cite{LIGOScientific:2019dag,Romero-Shaw:2019itr} for eccentric
\ac{bbh} searches).

The \ac{bns} signals GW170817 and GW190425 detected by the LIGO-Virgo
collaboration~\cite{TheLIGOScientific:2017qsa,Abbott:2017dke,Abbott:2020uma} have
been found and analyzed with quasicircular \ac{gw} templates. Their
eccentricities are (or are assumed to be) sufficiently small so that inference with
quasicircular templates does not affect the source parameters. 
An eccentric \ac{pe} of the two signals using a 
\ac{pn} Taylor F2 approximant indicates upper
limits of $e_0\leq0.024$ and $e_0\leq0.048$ for GW170817 and GW190925,
respectively, at 90\% confidence level \cite{Lenon:2020oza}.
The direct \ac{pe}
with eccentric templates increases the
upper limit by a factor of 3 when compared to estimates with more
approximated methods \cite{Romero-Shaw:2020aaj}.

Several recent studies have analyzed eccentricity in \ac{bbh}
observations~\cite{Gayathri:2020coq,Gayathri:2020fbl,Romero-Shaw:2021ual,OShea:2021ugg,Romero-Shaw:2020aaj,Gamba:2021gap,Romero-Shaw:2022xko,Iglesias:2022xfc,Bonino:2022hkj}.  
Some of them suggest that signals from highly eccentric mergers and/or
head-on collisions may be confused with mildly precessing
quasicircular binaries \cite{CalderonBustillo:2020odh,Romero-Shaw:2020thy,Gamba:2021gap}. This
degeneracy might be particularly relevant for events like GW190521 but
could in principle play a role also in the interpretation of binaries
with lower masses, including \acp{bns}. However, the systematics in \ac{gw}
\ac{pe} for the latter signals are not yet fully understood.

\citet{Favata:2021vhw} assessed the detectability and biases in \ac{bns} 
eccentricity measurement using spin-aligned \ac{pn} models and Fisher matrix. They find that a
measurement of eccentricity better than 30\% (one-sigma fractional error) requires
eccentricity values larger than 0.01 at 10~Hz. Low frequency sensitivity is
particularly relevant for eccentricity measurements. Systematic errors
become comparable to these statistical errors for eccentricities as
low as 0.01. One of the main reasons for this bias is a (Newtonian)
degeneracy with the chirp mass. 
\citet{Cho:2022cdy} performed Bayesian \acp{pe} of
nonspinning eccentric signal with $e_0\leq0.025$ using nonspinning
TaylorF2 quasicircular waveforms. They find the use of circular
templates introduces systematic biases in 
the intrinsic parameters (chirp mass, symmetric mass ratio and tidal
parameters) and calculate the distributions of the biases with
Markov-Chain-Monte-Carlo methods (for the considered range of eccentricities).

Compact binaries analyses of eccentric signals rely on the
availability of accurate waveform models. Several binary black
holes models are available~\cite{Huerta:2017kez, Cao:2017ndf, Chiaramello:2020ehz, Nagar:2021xnh, Yun:2021jnh, Ramos-Buades:2021adz, Nagar:2024dzj, Liu:2023ldr, Gamboa:2024hli, Planas:2025feq}
and have been adopted in searches and parameter estimation, \eg,~\cite{Romero-Shaw:2020thy,Gayathri:2020coq,Gamba:2021gap,Bonino:2022hkj, Iglesias:2022xfc,Ramos-Buades:2023yhy, LIGOScientific:2023lpe, Gadre:2024ndy, Bhaumik:2024cec}.
Among them, the \ac{eob} \TEOB{-\dali} can generate waveforms for generic compact binaries and arbitrary orbits~\cite{Gamba:2024cvy, Albanesi:2025txj}. \TEOB{-\dali} 
incorporates tidal effects and can thus generate \ac{bns} waveforms for eccentric and spin-precessing mergers.

In this work, we consider nonspinning unequal-mass \ac{bns} signals
up to eccentricities $e_0\sim0.1$ and perform
full Bayesian mock analyses using quasicircular templates with either nonprecessing or spin-precessing effects.
Focusing on a case study that has intrinsic parameters compatible with 
GW170817 (including the eccentricity),
we demonstrate the biases in the inference of source parameters caused by neglecting noncircular effects in the waveform model.

The paper is structured as follows. In Sec.~\ref{sec:met}, we introduce the framework employed 
for the simulation of the signals and the relative analyses. We present the results we 
obtained, with a comparison between nonprecessing and spin-precessing effects in Sec.~\ref{sec:res}. Finally 
in Sec.~\ref{sec:conc}, we summarize and conclude.
\\

\subsection{Conventions}
We use geometric units $c=G=1$ with masses expressed in solar masses $\Msun$ 
or explicitly state units.
The binary mass is $M= m_1 +m_2$, where $m_{1,2}$ are the masses of the two stars, the mass ratio $q = m_1 /m_2 \ge 1$,
and the symmetric mass ratio $\nu = m_1 m_2 / M^2$. The chirp mass is $\Mc = \nu^{3/5}M$.
The dimensionless spin vectors are denoted with $\vec{\chi}_i$ for $i=1,2$ and
the orbital angular momentum with $\vec{L}$. 

The spin parameters parallel and perpendicular to $\hat{n}=\vec{L}/||\vec{L}||$ are 
\be\label{eq:chieff}
\chieff := \frac{m_1}{M}\vec{\chi}_1\cdot\hat{n}\, + (1\leftrightarrow2)
\ee
and 
\be\label{eq:chip}
\chip := {\rm max}\left(\chi_{1,\perp}, 
\frac{1}{q}\,\frac{4+3q}{4q+3}\,\chi_{2,\perp}
\right)
\ee
where $\chi_{i,\perp} = |\vec{\chi}_i - (\vec{\chi}_i\cdot\hat{n})\hat{n}|$
for $i=1,2$.
The quadrupolar tidal polarizabilities are defined 
as $\Lambda_{i}=({2}/{3})\,k_{2,i}\,C_i^{-5}$ for $i=1,2$,
where $k_{2,i}$ and $C_i$ are, respectively, the $\ell=2$ gravitoelectric Love
number and the compactness of the $i$th \ac{ns}.
The reduced tidal parameter~\cite{Favata:2013rwa},
\be\label{eq:tLam}
\tLam :=\frac{16}{13}\left[\frac{(m_1+12m_2)m_1^4\Lambda_1}{M^5} + (1\leftrightarrow 2)\right]\,,
\ee
determines the strength of the leading-order tidal contribution to the binary interaction potential, while the parameter
\be
\delta\tLam := 
\left[1 -\frac{7996}{1319}  \frac{m_2}{m_1} -\frac{11005}{1319} \left(\frac{m_2}{m_1} \right)^2\right]\frac{m_1^6\Lambda_1}{M^6} - (1\leftrightarrow 2)\,,
\ee
determines the next-to-leading order contribution for asymmetric binaries.

The \ac{gw} polarizations $h_+$ and $h_\times$, plus and cross, respectively,
 are decomposed in $(\ell,m)$ multipoles as
\be
\label{eq:hdecomp}
h_+ - \i h_\times
=D_L^{-1}\sum_{\ell=2}^\infty\sum_{m=-\ell}^{\ell} h_{\ell m}(t)\,{}_{-2}Y_{\ell m}(\iota,\varphi),
\ee
where $D_L$ is the luminosity distance, 
${}_{-2}Y_{\ell m}$ are the $s=-2$
spin-weighted spherical harmonics
and $\iota,\varphi$ are, respectively, the polar and azimuthal 
angles that define the orientation of the binary with respect to the 
observer. 
Each mode $h_{\ell m}(t)$ is decomposed 
in amplitude $A_{\ell m}(t)$ and phase $\phi_{\ell m}(t)$, as
\be
\label{eq:hlm}
h_{\ell m}(t) = A_{\ell m}(t)\,\e^{- \i \phi_{\ell m}(t)} \,,
\ee
with a related \ac{gw} angular frequency, $\omega_{\ell m}(t) =2\pi f_{\ell m}(t) = \dot{\phi}_{\ell m}(t)$.
If the multipolar indices $(\ell, m)$ are omitted from a multipolar
quantity, we implicitly refer to the dominant $(2,2)$ mode.

The Fourier transform of a time-domain signal $h_\lm(t)$ is indicated
as $\tilde{h}_{\ell m}(f)$ and defined as
\be
\label{eq:fourier}
\tilde{h}_{\ell m}(f) = \int_{-\infty}^{+\infty} h_{\ell m}(t)\, \e^{-2\pi\i f t  }\,\d t \,.
\ee
Analogously to the time-domain case, 
the frequency series $\tilde{h}_{\ell m}(f)$ can be decomposed  
in amplitude and phase. 
The numbers quoted from the posterior probabilities refer to the
median of the distributions and the 90\% \acp{ci}.

\section{Method}
\label{sec:met}

\subsection{Effective-one-body waveforms}
\label{sec:met:eob}

\begin{figure*}[t]
  \centering 
  \includegraphics[width=0.9\textwidth]{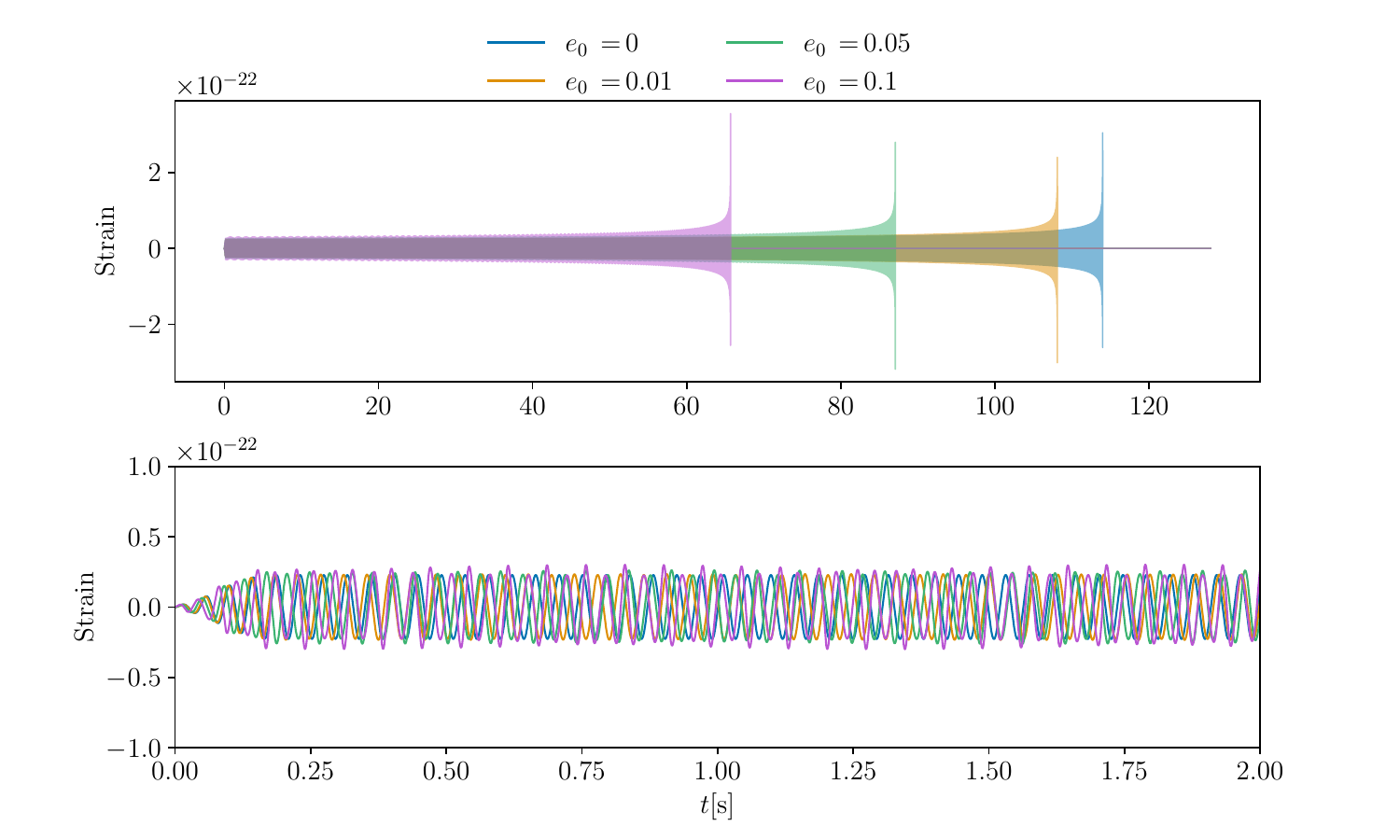}
  \caption{Injected waveforms for different 
  values of eccentricity, as reported in the legend.
  The injections have been performed with \TEOB{-\dali}~
  model, with $\Mc=1.1977\,\Msun$,  $q=1.5$, $\Lambda_1=400$, $\Lambda_2=600$, 
  $D_L=40\,{\rm Mpc}$, and $\iota = 45^\circ$.}
  \label{fig:wf}
\end{figure*}

We compute eccentric (and nonspinning) \ac{bns} signals using \TEOB{-\dali} \cite{Albanesi:2025txj} 
(see also \cite{Chiaramello:2020ehz, Gamba:2022mgx, Albanesi:2025txj} for details), 
a unified \ac{eob} approximant for the general relativistic dynamics and gravitational radiation of generic compact binaries.
When describing \acp{bns}, the conservative Hamiltonian of \TEOB{-\dali} incorporates point mass information up to 5\ac{pn}
\cite{Nagar:2018zoe}. Spin-orbit contributions are instead included at 
\ac{nnlo}~\cite{Damour:2014sva,Nagar:2021xnh} and inverse resummed following the \ac{eob} prescriptions detailed 
in Ref.~\cite{Nagar:2018zoe}. Finally, even-in-spin effects are accounted for up to \ac{nnlo} via the centrifugal radius.
The radiative sector contains orbital contributions at $3^{+2}$ \ac{pn} order--meaning that 3\ac{pn} terms
are integrated by 4PN and 5PN test-particle terms--and spin-spin interaction at \ac{nlo} for the $(2,2)$ mode
and \ac{lo} for the $(2,1)$,$(3,1)$, and $(3,3)$ modes.

Within \TEOB{-\dali}, the initial eccentricity $e_0$ and the initial orbit-averaged frequency $f_0$ determine
the initial conditions of the (Hamiltonian) equations of motion via
\begin{equation}
  r_0 = \frac{p_0}{1 + e_0\cos\zeta_0}\, ,
\end{equation}
where $r_0$ is the initial separation of the binary components, $p_0$ is the semilatus rectum of the orbit 
(which itself is a function of $f_0$ and $e_0$) 
and $\zeta_0$ is the initial true anomaly, in all cases considered set to $\pi$.
After initial conditions are determined, the \ac{eob} model evolves the system
under the influence of \ac{gw} radiation reaction, which includes noncircular
effects in its azimuthal component via the prescription described in Appendix~A of Ref.~\cite{Nagar:2024oyk},
and employs a nonzero radial component~\cite{Chiaramello:2020ehz}.

Inference (or recovery) is performed with the quasicircular (noneccentric) version of that model, 
which is consistent with \TEOB{-GIOTTO}~\cite{Nagar:2023zxh,Riemenschneider:2021ppj,Nagar:2019wds,Nagar:2018zoe},
and shares the same conservative dynamics and \acp{eom} as \TEOB{-\dali}.
The crucial difference is that no radial radiation reaction force and noncircular
effects are included in the description of the \ac{gw} flux and multipoles.
Both \ac{eob} models provide a faithful description of tidal effects up to merger \cite{Bernuzzi:2014owa,Akcay:2018yyh} 
and include spin precessional effects \cite{Gamba:2021ydi}.
For non eccentric binaries, \TEOB{} implements the postadiabatic approximation for the solution 
of the Hamilton equations~\cite{Nagar:2018gnk} and the EOB-SPA method to efficiently obtain 
frequency-domain waveforms~\cite{Gamba:2020ljo}. Both methods are key to speedup
the generation of long \ac{bns} waveforms.

\subsection{Bayesian parameter estimation}
\label{sec:met:pe}

The \ac{pe} experiments of this work are performed by injecting eccentric
\ac{bns} signals and recovering them with a full Bayesian analysis 
that employs the quasicircular model described
above. The Bayesian analyses are performed with the MPI-parallelized {\bajes}
pipeline~\cite{Breschi:2021wzr}. The Bayesian framework and the setup is summarized in what follows.
We refer to Ref.~\cite{Veitch:2009hd, LIGOScientific:2019hgc}
for a detailed discussion on Bayesian inference.

In the context of \ac{gw} data analysis,
the matched-filtered \ac{snr} of a template $h(t)$ with respect to
the data $d(t)$ is
\begin{equation}
\label{eq:snr}
	\rho:=\frac{(d|h)}{\sqrt{(h|h)}}\,,
\end{equation}
where the inner product 
between two time series $a(t)$ and $b(t)$ is defined in the frequency
space by
\begin{equation}
\label{eq:inner}
	(a|b) := \int \frac{\tilde a^*(f) \,  \tilde b(f)}{S_n(f)}\,\d f\,,
\end{equation}
with $S_n(f)$ the \ac{psd} of the noise segment.
For multiple detectors, Eq.~\eqref{eq:snr} can be generalized 
employing the sum in quadrature.\\
The inner product of Eq.~\eqref{eq:inner} forms also the basis for 
the likelihood function $p(d|h(\boldsymbol{\theta}),H)$,
\be
  \log p(d|h(\boldsymbol{\theta}),H) \propto -\dfrac{1}{2} \sum_i \big(d-h(\boldsymbol{\theta})\big|d-h(\boldsymbol{\theta})\big)_i\,,
\ee
where $H$ represents the signal hypothesis,
the subscript $i$ runs over the employed detectors 
and the parameter vector $\boldsymbol{\theta}$ corresponds to the combination 
of the \ac{bns} parameters we infer. Due to the complexity and the multidimensionality 
of the likelihood function, we employ nested sampling algorithms \cite{Veitch:2014wba,Handley:2015fda,Allison:2013npa}~to estimate the 
posterior $p(\boldsymbol{\theta}| d, H)$ and the evidence $Z = p(d,H)$, obtained by integrating the 
likelihood over the prior distribution of the model parameters.\\

The injected \ac{gw} signals correspond to
nonspinning eccentric 
\ac{bns} with $\Mc=1.1977\,\Msun$ (in the detector frame), 
$q=1.5$, $\Lambda_1=400$,
and $\Lambda_2=600$ ($\tLam=488$ and $\delta\tLam=-60$).
The considered eccentricity values are 
\be
e_0 = (0, 0.01, 0.05, 0.1)\,.
\ee
Mock \ac{gw} data are generated for the LIGO-Virgo three-detector network 
\cite{Aasi:2013wya,TheVirgo:2014hva,Abbott:2016blz,LIGOScientific:2018mvr,LIGOScientific:2019lzm}
with a duration of $128\,{\rm s}$ and a sampling rate of $4096\,$Hz.
Each signal is injected  with inclination angle $\iota=45^\circ$ at different distances, 
\be
D_L = (40,80,100)\,\Mpc.
\ee

The injections (see Fig.~\ref{fig:wf}) are performed with both zero noise or using Gaussian-and-stationary 
noise colored according to the \acp{psd} of the GW170817 event~\cite{LIGOScientific:2018mvr}.
The injected signals span the range of \acp{snr} $\rho_{\rm inj}\in[12,32]$.
The total number of mock \ac{gw} data (and thus \ac{pe} experiments) is $4\times3\times2=24$.

Our Bayesian \ac{pe} analyses employ the nested sampling implementation of ${\tt dynesty}$ 
\cite{Speagle:2019ivv} with a minimum of 3000 live points as baseline 
(used for most of the simulations). The likelihood is computed over the frequency range 
from $23\,$Hz to $2000\,$Hz  employing the GW170817 \acp{psd}~\cite{LIGOScientific:2018mvr}
and it is marginalized over reference time and phase.
Therefore, the sampling explores a 15-dimensional parameters space, 
\be
\params = \{m_1, m_2, \vec{\chi}_1, \vec{\chi}_2, \Lambda_1, \Lambda_2, D_L, \iota, \alpha, \delta, \psi\}\,,
\ee
where $(\alpha,\delta)$ identify the sky position of the source (right
ascension and declination, respectively) and 
$\psi\in[0,\pi]$ is the polarization angle. The prior
follows closely the standard choices described 
in~\cite{Breschi:2021wzr}
with isotropic spins
and volumetric luminosity distance.
The prior bounds are chosen as follows: 
\begin{subequations}
  \begin{align}
  &
  \Mc\in[1,1.5]\,,\ 
\\
  &
  q\leq q_{\rm max} = (4,10,50)\,,\
\\
&
|\vec{\chi}_i|\leq0.9\,,\ 
\\
&
\Lambda_i\in[0,5000]\,,\
\\
&
D_L\in[10,500]\Mpc\,.
\end{align}
\end{subequations}
The prior on the mass ratio has $q_{\rm max}=10$ for most of the
simulations but some simulations have considered the other values as
described below.\\
In the case of zero noise injections we performed the inference both using
aligned and precessing spins priors.

\section{Results}

\label{sec:res}

The results of our analysis are discussed in the following, starting
from the nonprecessing and moving to the precessing case. 
In either scenario we study the recovered \ac{snr} and source
parameters. In this section $e$ always refers to the eccentricity at $f_0=$ 23~Hz (\ie,~$e_0$).

\subsection{Nonprecessing inferences}
\label{sec:res:nonprec}

\begin{figure}[t]
  \centering 
  \includegraphics[width=0.45\textwidth]{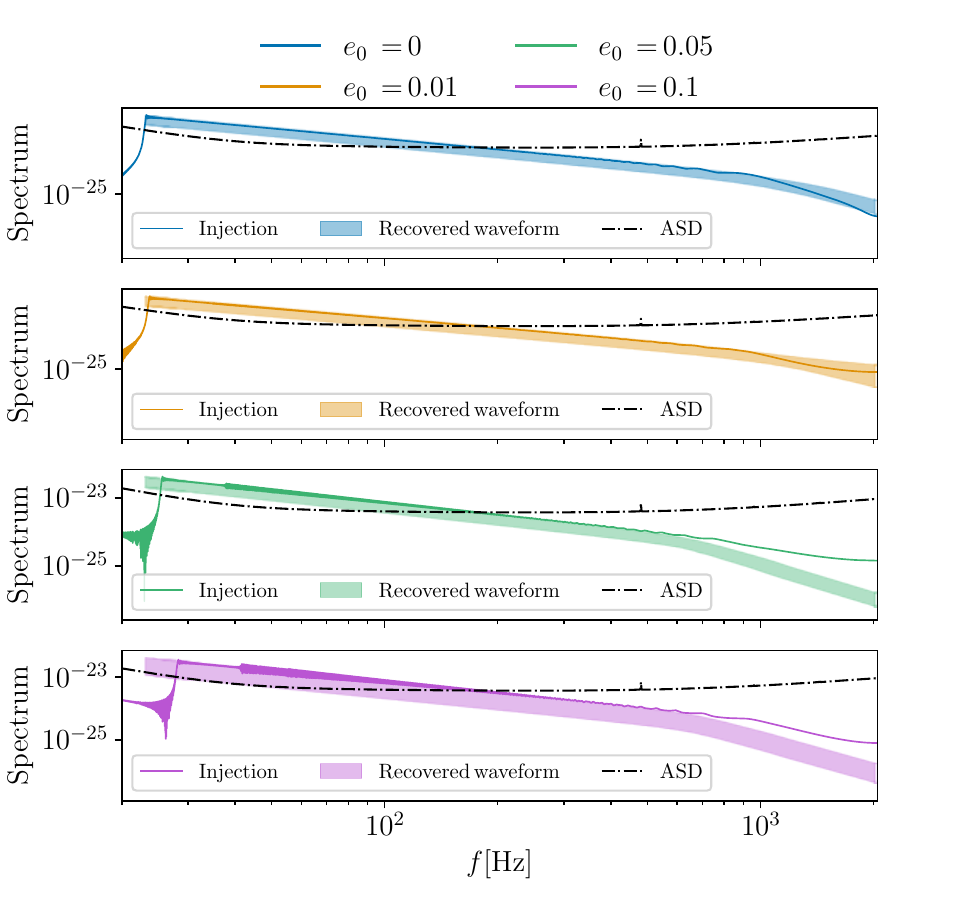}
  \caption{Injected and reconstructed waveforms in the spin-aligned case for different 
  values of eccentricity, as reported in the legend.
  The injections, plotted with solid lines, have been performed with \TEOB{-\dali}~
  model, with $\Mc=1.1977\,\Msun$,  $q=1.5$, $\Lambda_1=400$, $\Lambda_2=600$, 
  $D_L=40\,{\rm Mpc}$, and $\iota = 45^\circ$. The shaded regions are the $90\%$ credibility
  regions of the reconstructed spectrum and the dash-dotted line represents the amplitude 
  spectral density of the GW170817 event.}
  \label{fig:spec}
\end{figure}

In this subsection we discuss the results obtained when inference is performed with quasicircular nonprecessing waveforms. 
The results for both Gaussian and zero noise are collected in Table~\ref{tab:inject:alig}.

Examples of the recovered waveforms in frequency domain are shown in
Fig.~\ref{fig:spec}. The figure suggests that the most
informative portion of the signals is up ${\lesssim}200$~Hz~\cite{Gamba:2020wgg}. In this 
frequency range, the injected data (solid lines) appear well reproduced
by the best quasicircular waveform (colored region, corresponding to
90\% \ac{ci}); the agreement worsens for the largest simulated
eccentricities which are clearly not contained within the 90\% \acp{ci}.
At higher frequencies the merger waveform is captured only for the
smallest eccentricity and for the noneccentric binary, while the other two recoveries predict an earlier
merger than the injection. This qualitative analysis already suggests
that the chirp mass inferred from early frequencies overestimates the
injected one. Therefore, later in the text, we distinguish between low ($e \lesssim 0.01$) and 
high ($e \gtrsim 0.05$) eccentricity cases.

\begin{table*}[t]
  \centering    
  \caption{Summary table, spin-aligned \acp{pe}.}
  \resizebox{\textwidth}{!}{
    \begin{tabular}{cccccc|cccccc}        
      \hline
      \hline
      \multicolumn{6}{c|}{Injected properties}&\multicolumn{6}{c}{Recovered properties}\\
      \hline
      $e$ & $\Mecc$ & $D_L$ & $\rho_{\rm inj}$ & $q_{\rm max}$& Noise 
      & $\Mc$ & $q$ & $\chieff$ & $\tLam$ & $D_L$ & $\rho_{\rm rec}$ \\
       & $[\Msun]$ & $[\Mpc]$ & $$ & & $$ 
      & $[\Msun]$ & $ $ & $ $ & & $[\Mpc]$ & $ $ \\
      \hline
      \hline
      \multirow{3}*{0} & \multirow{3}*{1.1977 }& 40 & 31.7 & 4 &Z &$1.1977^{+0.0003}_{-0.0002}$ & $1.48^{+0.65}_{-0.41}$ & $0.003^{+0.066}_{-0.029}$ & $573^{+455}_{-319}$ & $36^{+19}_{-12}$ & $31.05^{+0.59}_{-0.57}$\\
      &  & 80 & 15.9 & 4 & Z & $1.1979^{+0.0006}_{-0.0003}$ & $1.52^{+1.05}_{-0.45}$ & $0.016^{+0.118}_{-0.038}$ & $1046^{+1849}_{-864}$ & $78^{+57}_{-32}$ & $15.38^{+0.32}_{-0.46}$\\
      &  & 100 & 12.7 & 4 & Z & $1.1979^{+0.0007}_{-0.0004}$ & $1.46^{+1.14}_{-0.40}$ & $0.015^{+0.134}_{-0.037}$ & $1443^{+2286}_{-1139}$ & $105^{+72}_{-52}$ & $12.21^{+0.33}_{-0.49}$\\
      \hline
      \multirow{3}*{0.01} & \multirow{3}*{1.982}& 40 & 31.7 & 10 & Z & $1.1980^{+0.0004}_{-0.0002}$ & $1.60^{+0.91}_{-0.50}$ & $0.023^{+0.091}_{-0.040}$ & $623^{+546}_{-414}$ & $37^{+19}_{-13}$ & $31.02^{+0.58}_{-0.59}$\\
      &  & 80 & 15.9 & 10 & Z & $1.1981^{+0.0006}_{-0.0003}$ & $1.53^{+1.09}_{-0.45}$ & $0.024^{+0.124}_{-0.037}$ & $1116^{+1838}_{-878}$ & $75^{+53}_{-33}$ & $15.37^{+0.32}_{-0.48}$\\
      &  & 100 & 12.7 & 10 & Z & $1.1982^{+0.0008}_{-0.0004}$ & $1.54^{+1.11}_{-0.46}$ & $0.032^{+0.153}_{-0.047}$ & $1520^{+2370}_{-1230}$ & $110^{+69}_{-54}$ & $12.21^{+0.31}_{-0.52}$\\
      \hline
      \multirow{3}*{0.05} & \multirow{3}*{1.2096}& 40 & 31.6 & 10 & Z & $1.2091^{+0.0013}_{-0.0014}$ & $9.23^{+0.69}_{-1.87}$ & $0.652^{+0.090}_{-0.080}$ & $120^{+98}_{-58}$ & $38^{+18}_{-13}$ & $28.99^{+0.45}_{-0.47}$\\
      &  & 80 & 15.8 & 10 & Z & $1.2076^{+0.0040}_{-0.0044}$ & $6.93^{+2.90}_{-5.31}$ & $0.604^{+0.206}_{-0.326}$ & $249^{+3936}_{-178}$ & $89^{+39}_{-47}$ & $14.28^{+0.28}_{-0.47}$\\
      &  & 100 & 12.6 & 10 & Z & $1.2057^{+0.0060}_{-0.0033}$ & $4.14^{+5.71}_{-2.81}$ & $0.522^{+0.287}_{-0.338}$ & $750^{+3947}_{-641}$ & $114^{+65}_{-56}$ & $11.30^{+0.31}_{-0.59}$\\
      \hline
      \multirow{3}*{0.1} & \multirow{3}*{1.2473}& 40 & 31.4 & 10 & Z & $1.2197^{+0.0018}_{-0.0013}$ & $9.24^{+0.70}_{-1.60}$ & $0.829^{+0.069}_{-0.050}$ & $94^{+52}_{-22}$ & $53^{+20}_{-20}$ & $24.12^{+0.47}_{-0.59}$\\
      &  & 80 & 15.7 & 10 & Z & $1.2127^{+0.0064}_{-0.0036}$ & $2.47^{+6.72}_{-1.35}$ & $0.587^{+0.225}_{-0.271}$ & $1015^{+3118}_{-950}$ & $117^{+66}_{-55}$ & $11.67^{+0.41}_{-0.64}$\\
      &  & 100 & 12.6 & 10 & Z & $1.2113^{+0.0066}_{-0.0037}$ & $1.75^{+5.41}_{-0.68}$ & $0.449^{+0.298}_{-0.218}$ & $1419^{+2776}_{-1361}$ & $160^{+103}_{-83}$ & $9.05^{+0.49}_{-0.71}$\\      
      \hline
      \hline
      \multirow{3}*{0} & \multirow{3}*{1.1977}& 40 & 33.9 & 4 & G & $1.1977^{+0.0002}_{-0.0001}$ & $1.40^{+0.55}_{-0.33}$ & $-0.007^{+0.054}_{-0.021}$ & $587^{+275}_{-301}$ & $36^{+16}_{-14}$ & $33.20^{+0.75}_{-0.78}$\\
      &  & 80 & 14.5 & 4 & G & $1.1980^{+0.0006}_{-0.0003}$ & $1.42^{+1.08}_{-0.37}$ & $0.013^{+0.121}_{-0.032}$ & $1256^{+2016}_{-894}$ & $83^{+48}_{-38}$ & $14.43^{+0.56}_{-0.55}$\\
      &  & 100 &  12.0 & 4 & G & $1.1981^{+0.0007}_{-0.0005}$ & $1.51^{+1.18}_{-0.44}$ & $0.026^{+0.137}_{-0.049}$ & $2397^{+1909}_{-1821}$ & $102^{+70}_{-65}$ & $12.32^{+0.36}_{-0.57}$\\
      \hline
      \multirow{3}*{0.01} & \multirow{3}*{1.1982}& 40 & 32.9 & 4 & G & $1.1980^{+0.0004}_{-0.0002}$ & $1.45^{+0.79}_{-0.39}$ & $0.009^{+0.086}_{-0.027}$ & $641^{+648}_{-313}$ & $40^{+16}_{-18}$ & $32.16^{+0.67}_{-0.64}$\\
      &  & 80 & 17.5 & 4 & G & $1.1981^{+0.0006}_{-0.0003}$ & $1.48^{+1.05}_{-0.41}$ & $0.019^{+0.130}_{-0.038}$ & $1101^{+2101}_{-884}$ & $66^{+43}_{-30}$ & $17.22^{+0.34}_{-0.47}$\\
      &  & 100 & 12.4 & 4 & G & $1.1983^{+0.0007}_{-0.0005}$ & $1.43^{+1.04}_{-0.38}$ & $0.030^{+0.131}_{-0.040}$ & $1472^{+1814}_{-1094}$ & $98^{+63}_{-52}$ & $12.20^{+0.47}_{-0.56}$\\
      \hline
      \multirow{3}*{0.05} & \multirow{3}*{1.2096}& 40 & 30.6 & 4 & G & $1.2063^{+0.0026}_{-0.0014}$ & $3.60^{+0.39}_{-1.43}$ & $0.625^{+0.267}_{-0.157}$ & $1600^{+2633}_{-555}$ & $49^{+13}_{-23}$ & $28.13^{+0.42}_{-0.47}$\\
      &  & 80 & 16.0 & 4 & G & $1.2044^{+0.0030}_{-0.0021}$ & $2.38^{+1.53}_{-1.20}$ & $0.387^{+0.265}_{-0.235}$ & $1733^{+2595}_{-1201}$ & $81^{+48}_{-37}$ & $14.83^{+0.35}_{-0.53}$\\
      &  & 100 & 13.5 & 4 & G & $1.2038^{+0.0035}_{-0.0016}$ & $2.05^{+1.83}_{-0.88}$ & $0.345^{+0.330}_{-0.203}$ & $2954^{+2434}_{-2310}$ & $94^{+66}_{-58}$ & $12.66^{+0.35}_{-0.47}$\\
      \hline
      \multirow{3}*{0.1} & \multirow{3}*{1.2473}& 40 & 31.6 & 10 & G & $1.2185^{+0.0008}_{-0.0010}$ & $8.67^{+1.14}_{-1.88}$ & $0.800^{+0.039}_{-0.039}$ & $104^{+77}_{-30}$ & $52^{+19}_{-20}$ & $25.21^{+0.50}_{-0.51}$\\
      &  & 80 & 14.5 & 10 & G & $1.2076^{+0.0113}_{-0.0013}$ & $1.97^{+7.29}_{-0.85}$ & $0.355^{+0.438}_{-0.162}$ & $2137^{+2378}_{-2059}$ & $108^{+63}_{-51}$ & $12.20^{+0.41}_{-0.55}$\\
      &  & 100 & 12.2 & 10 & G & $1.2091^{+0.0019}_{-0.0009}$ & $1.44^{+1.60}_{-0.40}$ & $0.318^{+0.207}_{-0.091}$ & $2374^{+1604}_{-1911}$ & $140^{+89}_{-66}$ & $9.79^{+0.44}_{-0.62}$\\
      \hline
      \hline
  \end{tabular}}
  \label{tab:inject:alig}
\end{table*}

\begin{figure}[t]
  \centering 
  \includegraphics[width=0.45\textwidth]{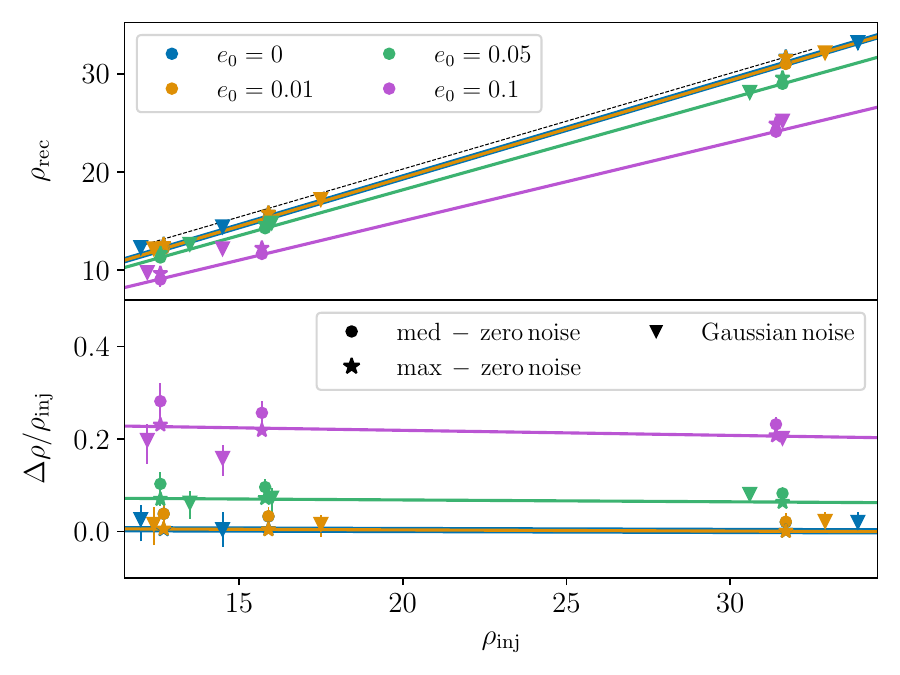}
  \caption{Recovered \ac{snr} (top panel) and \ac{snr} loss (bottom panel)
  				 as function of the injected \ac{snr} for spin-aligned \acp{pe}.
  				 Different colors refer to different eccentricities as reported in the legend.
  				 Round markers show the results of the median values of matched-filtered \ac{snr},
           star markers of the maximum values of the \ac{snr}, both for the 
           zero noise case, whereas triangular markers show the median values of the \ac{snr}
           in the case of Gaussian noise.}
  \label{fig:a_snr_loss}
\end{figure}

Figure~\ref{fig:a_snr_loss} summarizes the loss of matched-filtered \ac{snr} due to the
use of the quasicircular spin-aligned templates in the analyses of the
eccentric signals. 
We define the \ac{snr} loss $\Delta\rho= \rho_{\rm inj}-\rho_{\rm rec}$ 
as the difference between 
the \ac{snr} of the injected template $\rho_{\rm inj}$
and the \ac{snr} recovered with quasicircular template $\rho_{\rm rec}$.
The \ac{snr} loss increases with the
eccentricity, raising from ${\sim}2\%$ for $e=0.01$ 
to ${\sim}25\%$ for $e=0.1$.

If we consider the maximum value of the matched-filtered \ac{snr} (diamond markers) instead of the 
median one (round markers), the \ac{snr} loss decreases to ${\sim}0.5\%$ for $e=0.01$ 
to ${\sim}20\%$ for $e=0.1$. This is due to the fact that, in the case of zero noise,
the maximum value of the \ac{snr} achievable is the injected one, since there is no noise that 
can shift the peak of the distribution to larger values. 
Furthermore, using the maximum values of the \ac{snr},
the relative \ac{snr} loss $\Delta\rho/\rho_{\rm inj}$ is almost independent
on the injected \ac{snr} $\rho_{\rm inj}$. 
Hence, for the considered \ac{snr} and eccentricity
ranges, 
we find that the \ac{snr} loss can be fit to 
\be
\label{eq:snr-loss-fit}
\frac{\Delta\rho}{\rho_{\rm inj}} \approx \alpha  e\,,
\ee
with $\alpha=2.27\pm 0.17$. \footnote{In order to reduce the impact of noise 
fluctuations, the presented relation is calibrated 
only on zero noise results.}
Notably, these results are rather independent on the noise
realization for $e\lesssim 0.05$, as the variations of $\Delta\rho/\rho_{\rm inj}$
for the Gaussian noise simulations (triangular markers) are in agreement with the 
fits for zero noise. 

\begin{figure*}[t]
  \centering 
  \includegraphics[width=0.45\textwidth]{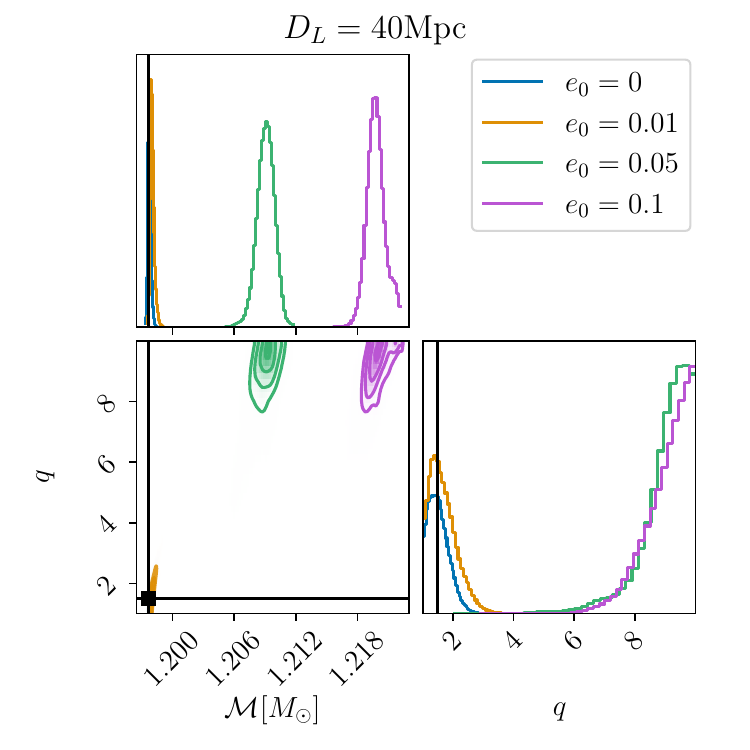}
  \includegraphics[width=0.45\textwidth]{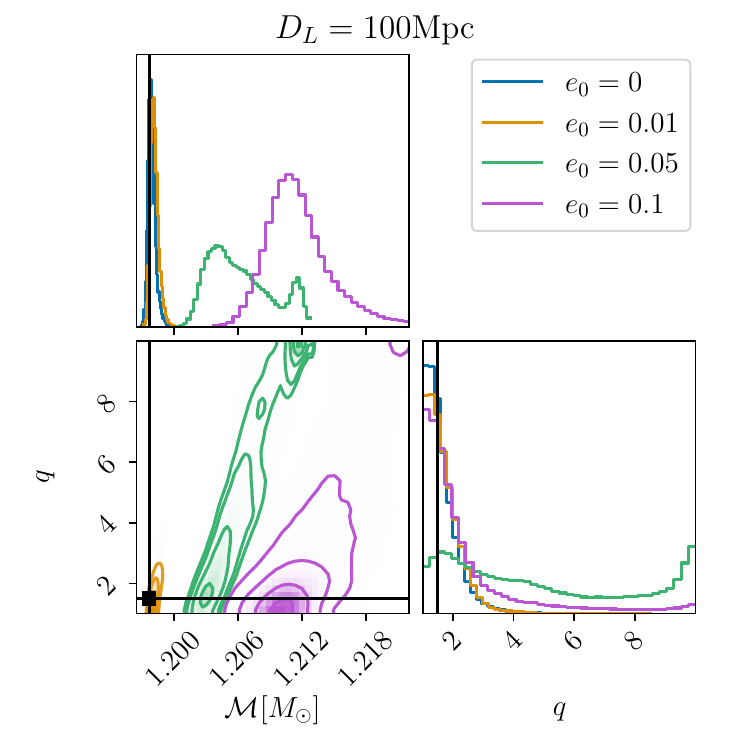}
  \caption{Posterior distribution of chirp masses and mass ratios for spin-aligned \acp{pe} with zero 
  noise, for two different luminosity distances $D_L=40\,{\rm Mpc}$ ($\rho_{\rm inj} = 31.7$) and 
  $D_L=100\,{\rm Mpc}$ ($\rho_{\rm inj} = 12.7$). Injected values are identified by the black lines.}
  \label{fig:corner:a_Mc}
\end{figure*}

\begin{figure}[t]
  \centering 
  \includegraphics[width=0.45\textwidth]{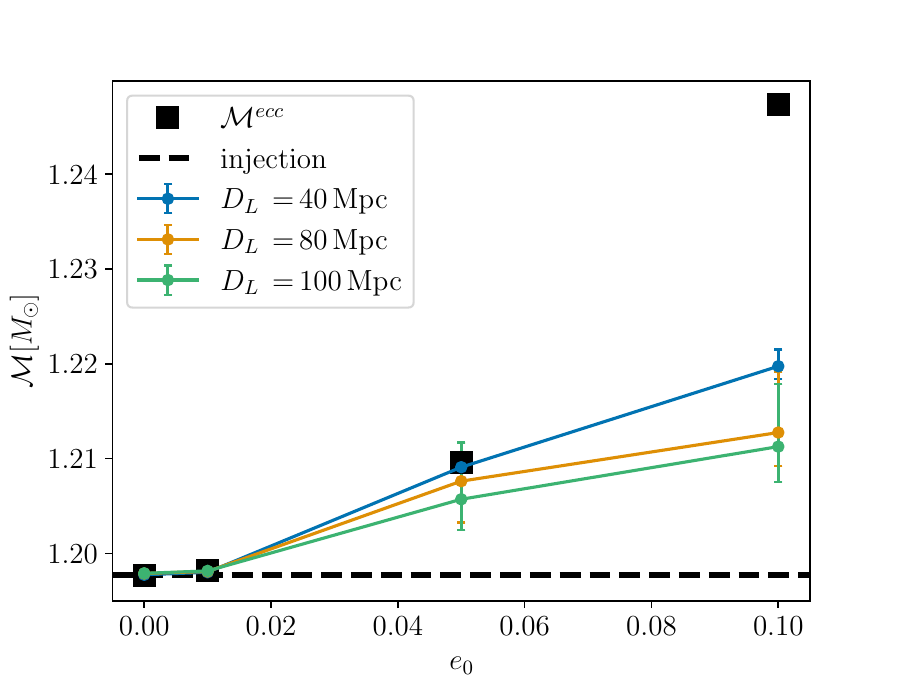}
  \caption{Recovered $\Mc$ as function of the eccentricity for spin-aligned \acp{pe}.
           Different colors refer to different injection distances as reported in the legend.
           The black dots show the leading order approximation for $\Mc_{\rm ecc}$ as per 
           Eq.~\eqref{eq:mchirp-ecc} }
  \label{fig:a_rec_Mecc}
\end{figure}

\subsubsection{Low eccentricity ($e\lesssim 0.01$)}
\label{sec:res:nonprec:small}

Figure~\ref{fig:corner:a_Mc} shows the recovered chirp
masses and mass ratios for the injections at $D_L = 40,\,100\,{\rm Mpc}$ (zero noise).
In both zero and low eccentricity cases, the posteriors are consistent 
with the injected data. However, for $e = 0.01$, we note a shift to larger 
median value. This is also confirmed in Fig.~\ref{fig:a_rec_Mecc} and Table~\ref{tab:inject:alig}.

The bias in the eccentric scenario can be interpreted with a Newtonian argument~\cite{Favata:2021vhw}. 
Chirp mass and eccentricity enter the leading order expression of the waveform's 
phase via the combination
\be
\label{eq:mchirp-ecc}
\Mecc = \Mc \left(1-\frac{157}{24} e^2\right)^{-3/5}\,.
\ee
The above expression is valid for small eccentricities and low
frequencies.  
Since the quasicircular templates have $e=0$, the eccentric signal can
only be reproduced with a larger $\Mc$.

\begin{figure}[t]
  \centering 
  \includegraphics[width=0.45\textwidth]{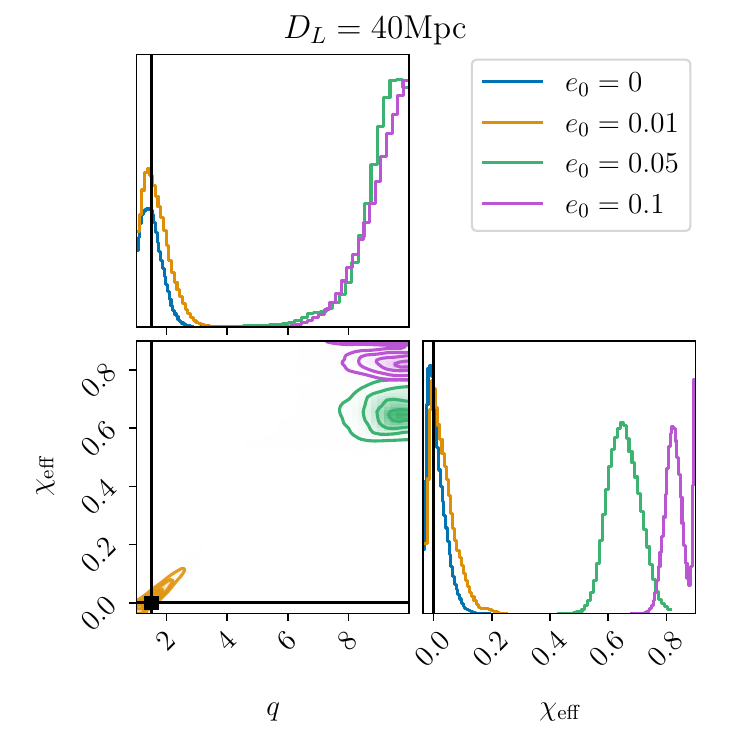}
    \caption{Posterior distribution of effective spin parameters and mass ratios for spin-aligned 
  \acp{pe} with zero noise for $D_L=40\,{\rm Mpc}$ ($\rho_{\rm inj} = 31.7$). 
  Injected values are identified by the black lines.}
  \label{fig:corner:a_spinq}
\end{figure}

Figure~\ref{fig:corner:a_spinq} shows the posterior distributions of mass ratios and 
effective spin $\chieff$ for $D_L = 40\,{\rm Mpc}$.
They are both compatible with the injected values in the non eccentric configuration 
as well as for $e=0.01$ (see Table~\ref{tab:inject:alig}).

\begin{figure}[t]
  \centering 
  \includegraphics[width=0.45\textwidth]{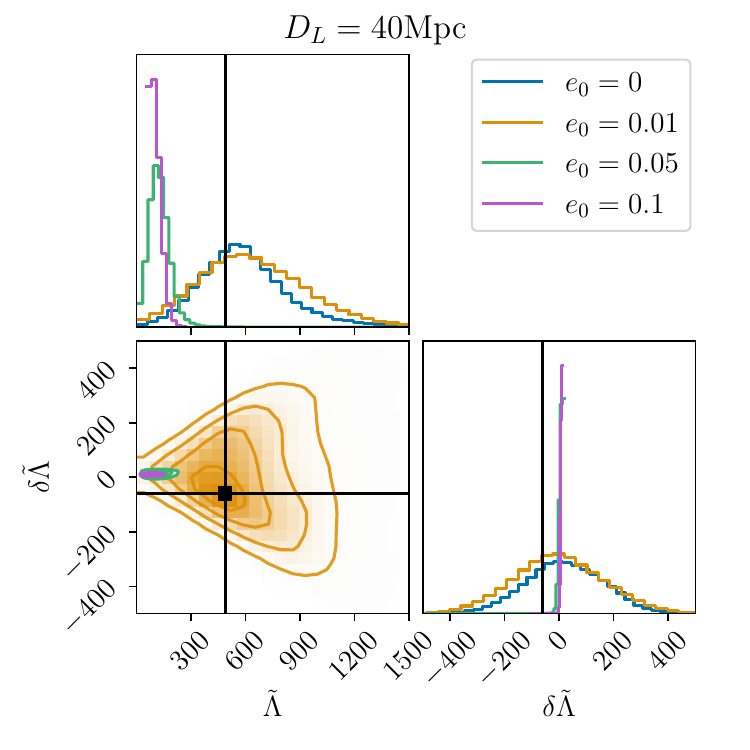}
   \caption{Posterior distributions of tidal polarizability parameters and asymmetric 
   tidal parameters for spin-aligned \acp{pe} with 
   zero noise for $D_L=40\,{\rm Mpc}$ ($\rho_{\rm inj} = 31.7$). 
   Injected values are identified by the black lines.}
  \label{fig:corner:a_lambda}
\end{figure}

The same behavior is observed for the tidal parameters, plotted in Fig.~\ref{fig:corner:a_lambda} for $D_L=40\,\Mpc$.
The posteriors capture the correct value of $\tLam$ within the
statistical accuracy expected at these \ac{snr} for sufficiently small
eccentricities $e\sim0.01$. For comparison, the eccentricities
inferred in \cite{Lenon:2020oza} for GW170817 are a factor 2 to 4
larger. 

\subsubsection{High eccentricity ($e\gtrsim0.05$)}
\label{sec:res:nonprec:large}

In the analysis of high eccentric configurations, the most prominent feature of the posteriors 
is the bias in the chirp mass parameter. The posteriors of the chirp mass shown in 
Fig.~\ref{fig:corner:a_Mc}, are systematically shifted to larger 
median values as the eccentricity increases. 

Figure~\ref{fig:a_rec_Mecc} and Table~\ref{tab:inject:alig} demonstrate that the recovered $\Mc$ 
are consistent with the values predicted by Eq.~\eqref{eq:mchirp-ecc} 
within the 90\% confidence level up to eccentricity $e=0.05$ in the case of zero noise.
For the $e=0.1$ cases, the recovered values are smaller than the prediction 
of Eq.~\eqref{eq:mchirp-ecc}. This means that
Eq.~\eqref{eq:mchirp-ecc} is not accurate to quantitatively predict the
biases for eccentricities $e>0.05$. 

From Fig.~\ref{fig:corner:a_Mc} we can also comment about the mass ratio. 
The mass ratio inference is found significantly biased and strongly
degenerate with the chirp mass for $e\gtrsim0.05$. In all the analyses
with $e\gtrsim0.05$ and $q_{\rm max} = 4$ or $q_{\rm max}=10$ the mass ratio 
posteriors systematically rail against the $q_{\rm max}$ prior and, 
for the smallest luminosity distance considered, the posteriors are
sufficiently narrow to not support the injected value anymore.

\begin{figure}[t]
  \centering 
  \includegraphics[width=0.45\textwidth]{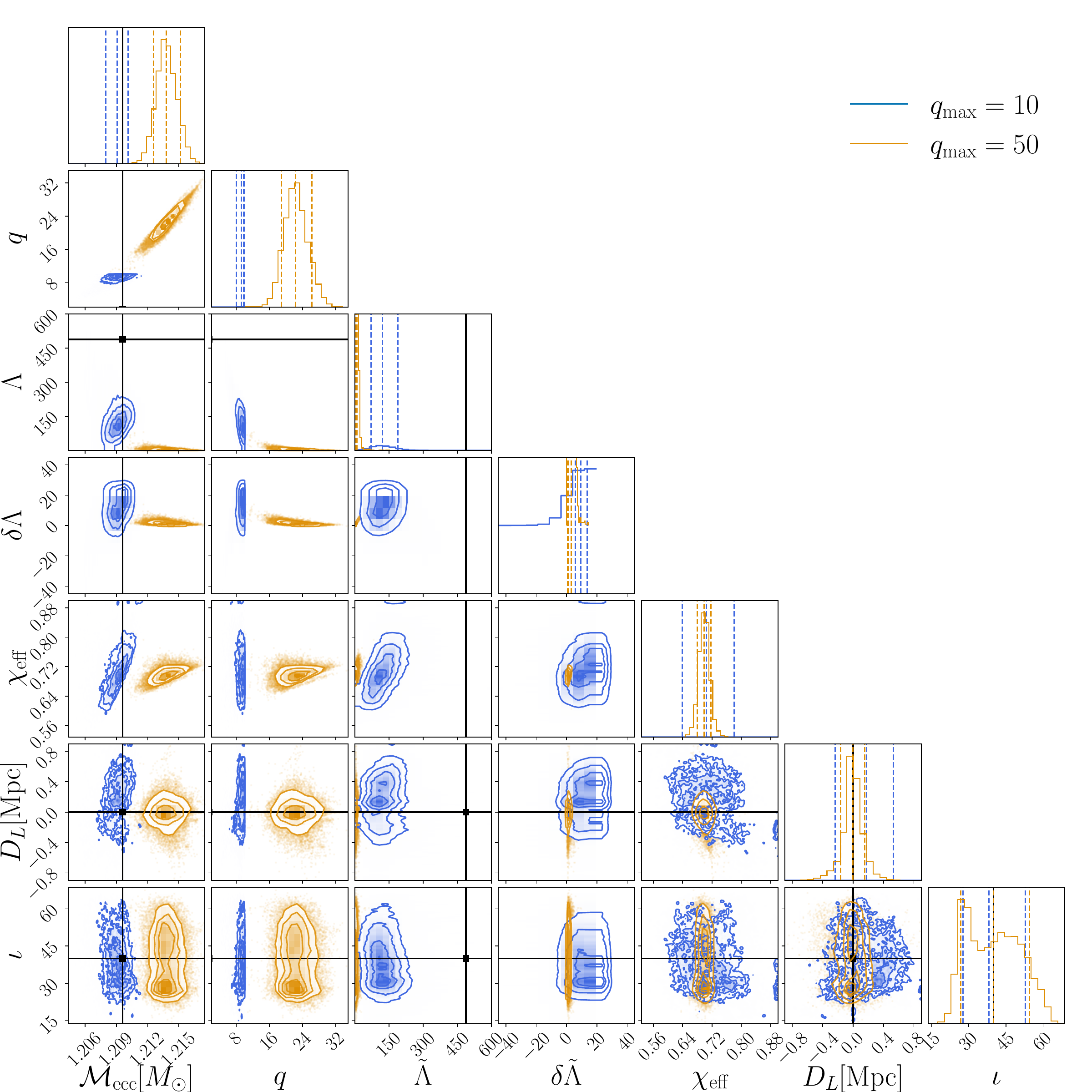}
  \caption{Corner plot for the $e = 0.05$, $D_L = 40\,\Mpc$ ($\rho_{\rm inj} = 31.6$) 
    spin-aligned configuration with an upper bound for the mass ratio of $q_{\rm
    max}=50$ (yellow) and $q_{\rm max}=10$ (blue). The black lines mark
    the injected values.} 
  \label{fig:corner:qmqx50_vs_10}
\end{figure}

This effect is further investigated in the case of $e = 0.05$ and $D_L = 40\,{\rm Mpc}$
by means of
(i) a \ac{pe} run with $q_{\rm max} = 10$ with increasing the live point of
the sampler to 5000, in order to test convergence;
and
(ii) a \ac{pe} run with $q_{\rm max} = 50$ (3000 live points). 
The first simulation shows again that
the $q$ posterior is railing against its upper bound. This suggests that
this effect is not due to the sampler.
The second simulation returns a converged posterior peaked
at $q = 22 \pm 5$. The posteriors of this run are shown in Fig.~\ref{fig:corner:qmqx50_vs_10}. 
This run also yields a chirp mass larger than 
$\Mecc$ from Eq.~\eqref{eq:mchirp-ecc}. In a real \ac{pe} the
single object masses would be inferred as $m_1\approx8\,\Msun$ and
$m_2\approx0.35\,\Msun$, neither of which can be identified as a \ac{ns}
object. 
The reduced tidal parameter inferred by this run is
$\tLam = 4.00^{+3.99}_{-1.81}$ which, together with the single object masses, yields
$\Lambda_1 = 2.57^{+2.52}_{-1.16}$ and $\Lambda_2=1553^{+ 2256}_{-988}$.
While the heavier binary component would be misidentified
as a \ac{bh} the secondary would be identified neither as a \ac{ns}
nor a \ac{bh}.

From Fig.~\ref{fig:corner:a_spinq} we note that significant biases are present already
at $e=0.05$ for the recovery of the effective spin. Similar to the chirp mass, 
the effective spin parameter is
progressively more biased for signal from higher eccentricities and at
close distances. 
All the inferences of eccentric \acp{bns} deliver a positive $\chi_{\rm eff}$.
Since the template we are using for the analysis does not include eccentricity, 
the \ac{pe} returns a larger value of $\chieff$ to compensate the eccentric effects 
in the form of spin-orbit interactions. 

Finally, Fig.~\ref{fig:corner:a_lambda} shows the posteriors of the
tidal parameters. At higher eccentricities $e\gtrsim 0.05$ the reduced tidal parameter is
biased toward significantly smaller values. This parameter is also
degenerate with the mass ratio: large values of $q$ usually move the
peak of the reduced tidal parameters posterior close to zero for small distances.
Combined with the bias in the chirp mass and mass ratio, this can lead
to an incorrect (and puzzling)  interpretation of the source as
discussed above. 

In summary, for high eccentric cases, we identify significant biases in the 
recovery of the source parameters employing quasicircular templates. This highlights
the inadequacy of quasicircular waveform models for analyzing eccentric signals.

\subsubsection{Gaussian noise}

The \acp{pe} relative to injections with Gaussian noise are generally consistent with 
the zero noise ones, as reported in Table~\ref{tab:inject:alig}. 
For $e=0$ and $e=0.01$, all the recovered values are coherent with our injections, except 
the chirp mass in the eccentric case, since the eccentricity is taken into account 
in $\Mecc$ [Eq.~\eqref{eq:mchirp-ecc}]. 
At higher eccentricities $e\gtrsim0.05$ the mass ratio is railing against the prior
and the chirp mass is larger than the injected value, but smaller than
what is expected from Eq.~\eqref{eq:mchirp-ecc}. The effective spins are 
not compatible with the nonspinning injected values, as a compensation of the 
eccentricity, which is not considered in our recovery model with the quasicircular 
template. As the eccentricity increases, the values of the tidal polarizability 
parameters are incompatible with the injected ones and we notice a loss in the \ac{snr}, 
similar to the zero noise case.

\subsection{Spin precession}
\label{sec:res:prec}

\begin{table*}[t]
  \centering    
  \caption{Summary table, spin-precessing \acp{pe}}
  \resizebox{\textwidth}{!}{
    \begin{tabular}{cccccc|ccccccc}        
      \hline
      \hline
      \multicolumn{6}{c|}{Injected properties}&\multicolumn{7}{c}{Recovered properties}\\
      \hline
      $e$ & $\Mecc$ & $D_L$ & $\rho_{\rm inj}$ & $q_{\rm max}$& Noise 
      & $\Mc$ & $q$ & $\chieff$ & $\chip$ & $\tLam$ & $D_L$ & $\rho_{\rm rec}$ \\
       & $[\Msun]$ & $[\Mpc]$ & $ $ & & $ $ 
      & $[\Msun]$ & $ $ & $ $ & & & $[\Mpc]$ & $ $ \\
      \hline
      \hline
      \multirow{3}*{0} & \multirow{3}*{1.1977 }& 40 & 31.7 & 4 & Z & $1.1978^{+0.0004}_{-0.0002}$ & $1.59^{+0.82}_{-0.49}$ & $0.014^{+0.083}_{-0.039}$ & $0.211^{+0.440}_{-0.183}$ & $546^{+538}_{-389}$ & $40^{+16}_{-15}$ & $31.11^{+0.52}_{-0.62}$ \\
      &  & 80 & 15.9 & 4 & Z & $1.1979^{+0.0005}_{-0.0003}$ & $1.53^{+1.04}_{-0.48}$ & $0.013^{+0.107}_{-0.039}$ & $0.286^{+0.432}_{-0.230}$ & $1091^{+1980}_{-872}$ & $82^{+43}_{-37}$ & $15.37^{+0.34}_{-0.47}$ \\
      &  & 100 & 12.7 & 4 & Z & $1.1979^{+0.0006}_{-0.0004}$ & $1.56^{+1.12}_{-0.50}$ & $0.022^{+0.117}_{-0.046}$ & $0.325^{+0.453}_{-0.267}$ & $1521^{+2253}_{-1250}$ & $105^{+65}_{-48}$ & $12.19^{+0.30}_{-0.51}$ \\
      \hline
      \multirow{3}*{0.01} & \multirow{3}*{1.1982}& 40 & 31.7 & 10 & Z & $1.1981^{+0.0004}_{-0.0002}$ & $1.78^{+0.70}_{-0.65}$ & $0.037^{+0.082}_{-0.052}$ & $0.308^{+0.331}_{-0.226}$ & $604^{+645}_{-382}$ & $41^{+13}_{-16}$ & $30.99^{+0.59}_{-0.63}$ \\
      &  & 80 & 15.8 & 10 & Z & $1.1981^{+0.0006}_{-0.0004}$ & $1.56^{+1.26}_{-0.49}$ & $0.027^{+0.123}_{-0.044}$ & $0.296^{+0.419}_{-0.228}$ & $1140^{+2121}_{-920}$ & $81^{+43}_{-33}$ & $15.34^{+0.32}_{-0.47}$ \\
      &  & 100 & 12.7 & 10 & Z & $1.1981^{+0.0007}_{-0.0004}$ & $1.52^{+1.24}_{-0.47}$ & $0.027^{+0.122}_{-0.045}$ & $0.312^{+0.433}_{-0.252}$ & $1769^{+2354}_{-1449}$ & $103^{+64}_{-44}$ & $12.18^{+0.30}_{-0.46}$ \\
      \hline
      \multirow{3}*{0.05} & \multirow{3}*{1.2096}& 40 & 31.6 & 10 & Z & $1.2093^{+0.0012}_{-0.0017}$ & $9.00^{+0.88}_{-2.08}$ & $0.668^{+0.079}_{-0.098}$ & $0.211^{+0.152}_{-0.138}$ & $141^{+132}_{-70}$ & $44^{+13}_{-15}$ & $29.02^{+0.48}_{-0.49}$ \\
      &  & 80 & 15.8 & 10 & Z & $1.2071^{+0.0035}_{-0.0041}$ & $5.73^{+4.07}_{-4.21}$ & $0.579^{+0.168}_{-0.343}$ & $0.297^{+0.303}_{-0.203}$ & $383^{+4362}_{-304}$ & $92^{+37}_{-38}$ & $14.31^{+0.28}_{-0.51}$ \\
      &  & 100 & 12.6 & 10 & Z & $1.2042^{+0.0056}_{-0.0021}$ & $2.31^{+6.94}_{-1.14}$ & $0.376^{+0.319}_{-0.233}$ & $0.375^{+0.331}_{-0.251}$ & $2918^{+2385}_{-2790}$ & $120^{+69}_{-57}$ & $11.29^{+0.30}_{-0.55}$ \\
      \hline
      \multirow{3}*{0.1} & \multirow{3}*{1.2473}& 40 & 31.7 & 10 & Z & $1.2194^{+0.0013}_{-0.0017}$ & $9.11^{+0.83}_{-2.41}$ & $0.817^{+0.044}_{-0.051}$ & $0.196^{+0.149}_{-0.132}$ & $95^{+95}_{-21}$ & $49^{+25}_{-22}$ & $24.06^{+0.49}_{-0.55}$ \\
      &  & 80 & 15.7 & 10 & Z & $1.2119^{+0.0058}_{-0.0029}$ & $1.60^{+6.85}_{-0.54}$ & $0.504^{+0.248}_{-0.211}$ & $0.412^{+0.287}_{-0.270}$ & $2643^{+1814}_{-2567}$ & $114^{+77}_{-56}$ & $11.69^{+0.38}_{-0.59}$ \\
      &  & 100 & 12.6 & 10 & Z & $1.2113^{+0.0066}_{-0.0037}$ & $1.75^{+5.41}_{-0.68}$ & $0.449^{+0.298}_{-0.218}$ & $0.421^{+0.286}_{-0.277}$ & $1419^{+2776}_{-1361}$ & $160^{+103}_{-83}$ & $9.12^{+0.43}_{-0.64}$ \\
      \hline
      \hline
  \end{tabular}}
  \label{tab:inject:prec}
\end{table*}

\begin{figure*}[ht]
  \centering 
  \includegraphics[width=0.45\textwidth]{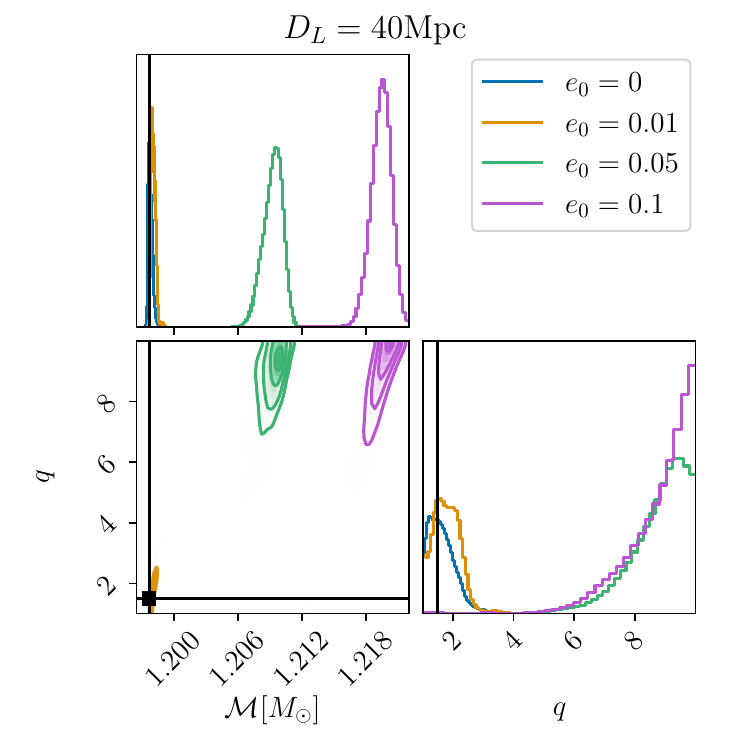}
  \includegraphics[width=0.45\textwidth]{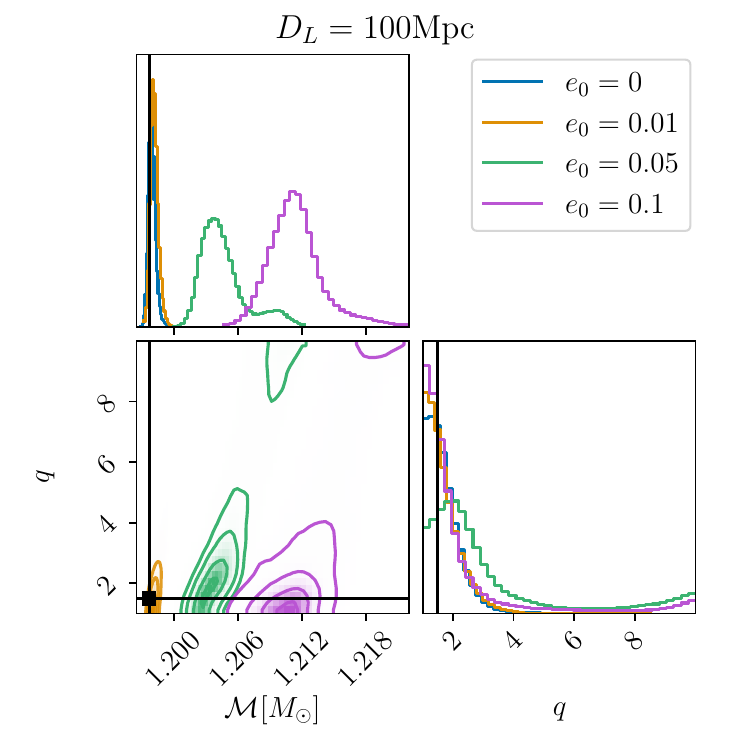}
  \caption{Posterior distributions of chirp masses and mass ratios for spin-precessing \acp{pe}
  for two different luminosity distances $D_L=40\,{\rm Mpc}$ ($\rho_{\rm inj} = 31.7$) and 
  $D_L=100\,{\rm Mpc}$ ($\rho_{\rm inj} = 12.7$). 
  Injected values are identified by the black lines.}
  \label{fig:corner:p_Mc}
\end{figure*}

\begin{figure*}[ht]
  \centering 
  \includegraphics[width=0.45\textwidth]{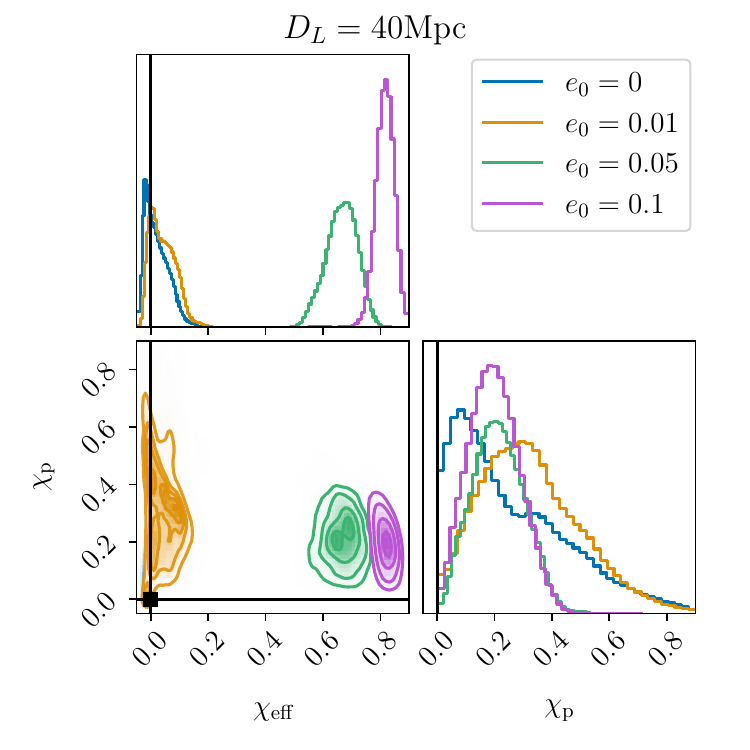}
  \includegraphics[width=0.45\textwidth]{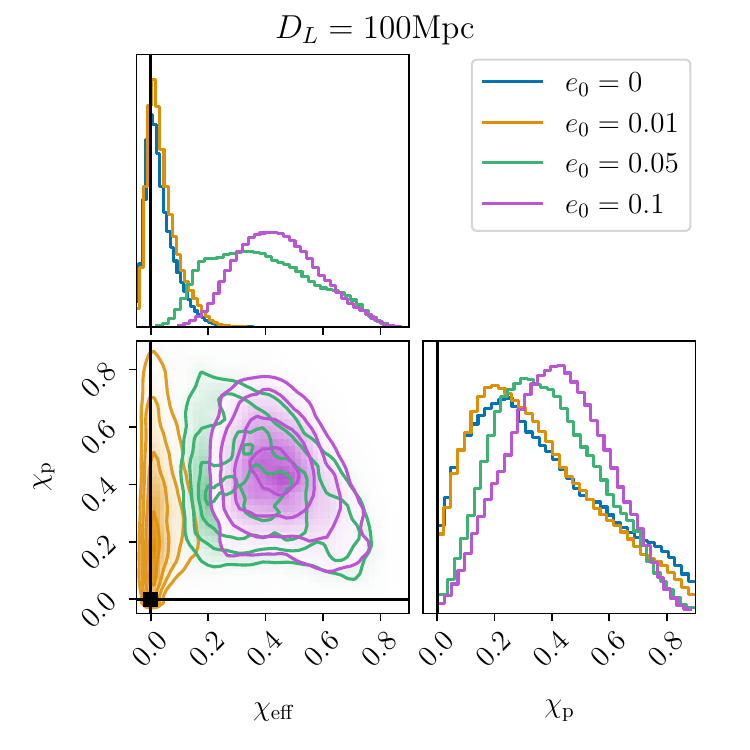}
  \caption{Posterior distributions of the parallel effective and perpendicular spins 
  for spin-precessing \acp{pe} for two different luminosity distances $D_L=40\,{\rm Mpc}$ 
  ($\rho_{\rm inj} = 31.7$) and $D_L=100\,{\rm Mpc}$ ($\rho_{\rm inj} = 12.7$). 
  Injected values are identified by the black lines.}
  \label{fig:corner:p_spin}
\end{figure*}

\begin{figure}[h]
  \centering 
  \includegraphics[width=0.45\textwidth]{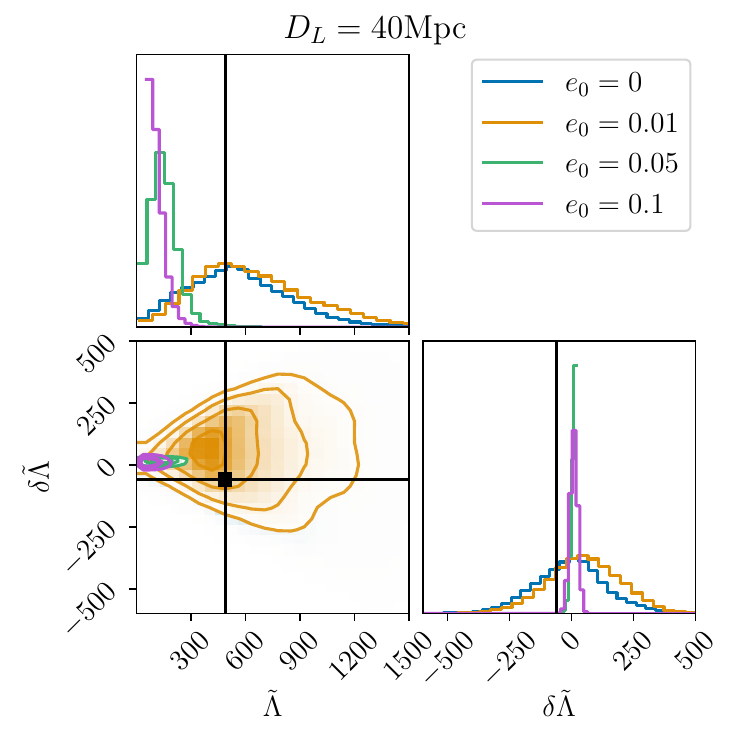}
    \caption{Posterior distributions of tidal polarizability parameters and asymmetric tidal 
  parameters for spin-precessing \acp{pe} for $D_L=40\,{\rm Mpc}$ ($\rho_{\rm inj} = 31.7$).
  Injected values are identified by the black lines.}
  \label{fig:corner:p_lambda}
\end{figure}

\begin{figure}[h]
  \centering 
  \includegraphics[width=0.45\textwidth]{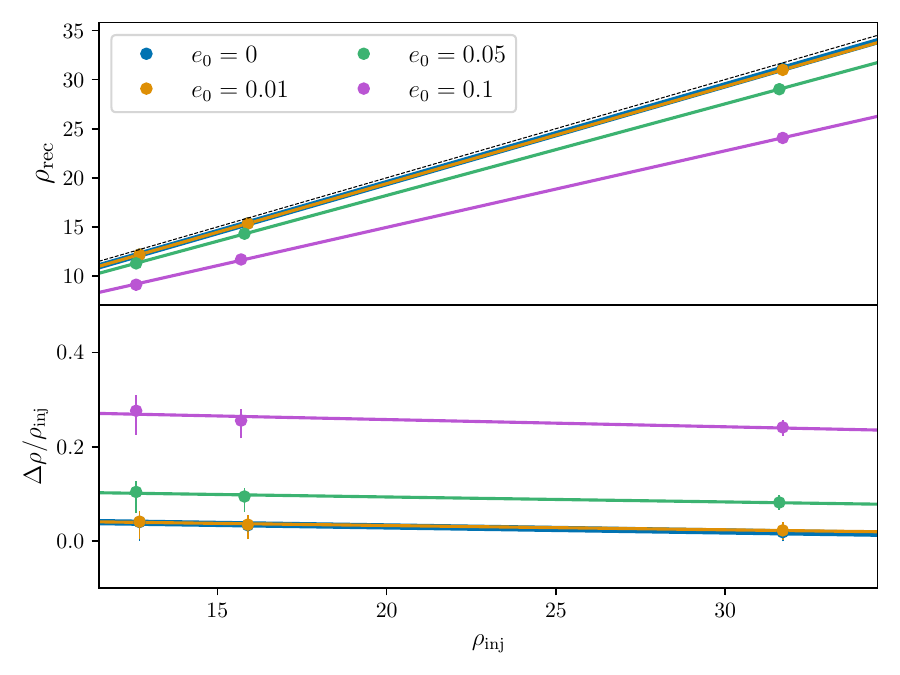}
  \caption{Median values of recovered \ac{snr} (top panel) and \ac{snr} loss (bottom panel)
           as function of the injected \ac{snr} for spin-precessing \acp{pe}.
           Different colors refer to different eccentricities as reported in the legend.
                                 }
  \label{fig:corner:p_snr_loss}
\end{figure}

In this subsection we discuss the results obtained when inference is performed with quasicircular precessing waveforms. 
The results are collected in Table~\ref{tab:inject:prec}. Since in the previous 
subsection we observed that the Gaussian noise cases are similar to the zero noise ones, we 
present here only the zero noise \acp{pe}. 

In the cases of zero and low eccentricity, the posterior distributions are compatible with the 
injected values for all the parameters. As the eccentricity increases, instead, we start observing 
the failure of the quasicircular template in recovering eccentric signals. 

Figure~\ref{fig:corner:p_Mc} shows the posterior distributions of chirp masses and mass ratios
for $D_L = 40,\,100\,{\rm Mpc}$. As in the aligned case we note that the chirp mass 
posteriors are shifted to larger values as the eccentricity increases and the values are compatible
with $\Mecc$ [Eq.\eqref{eq:mchirp-ecc}] for $e \lesssim 0.05$. 

The mass ratio has again the most significant bias for the eccentric \acp{bns}, as shown 
in Fig.~\ref{fig:corner:p_Mc}. For the smallest luminosity distance ($D_L = 40\,{\rm Mpc}$)
and $e\gtrsim0.05$, we obtain values of the mass ratio $\gtrsim9$. For the largest luminosity distance, 
\ie,~the smaller \ac{snr}, the injected values are instead compatible with the 
recovered ones within the $90\%$ \ac{ci}, meaning that there is a threshold \ac{snr} from 
which we start observing the biases.

The effective spin is progressively biased for higher eccentricities, as for the 
spin-aligned case. The spin perpendicular parameter, whose posterior distributions are
reported in Fig.~\ref{fig:corner:p_spin} together with the effective one at $D_L = 40,\,100\,{\rm Mpc}$,
is not well constrained.
We stress again that the template we are using for the inference is quasicircular, and
therefore the influence of the eccentricity is taken into account as spin-orbit effects,
leading to the increase of the effective spin for $e\gtrsim0.05$. 

Figure~\ref{fig:corner:p_lambda} shows the posterior distributions of the tidal polarizability 
parameters for $D_L = 40\,{\rm Mpc}$, again the result is in accordance with the spin-aligned 
case: at higher eccentricities $\tilde{\Lambda}$ is underestimated.

In Fig.~\ref{fig:corner:p_snr_loss} we report the \ac{snr} loss for the quasicircular spin-precessing
analysis. Similar to the spin-aligned case, as the eccentricity increases, the difference between the
injected and the recovered \ac{snr} increases, for $e = 0.01$ is $\sim 2\%$, whereas for $e=0.1$, it reaches
$\sim 25\%$. Since the results are comparable with the case of spin-aligned \acp{pe}, we can conclude 
that the eccentricity is not mimicked by the precession.

\section{Conclusions}
\label{sec:conc}

In this paper we demonstrated that use of quasicircular and spinning \ac{bns} waveforms in Bayesian inference 
of eccentric \ac{bns} signals with $e_0\gtrsim0.05$ can lead to significant parameter biases already at \ac{snr} 
${\gtrsim}12$. In our experiments several key parameters are profoundly affected by the use (assumption) of a 
noneccentric waveform. Biases in the chirp mass and mass ratio can be so severe to affect the source interpretation: 
in the worst case the \ac{bns} signal would be associated to a binary composed of a ${\sim}8\Mo$ black hole and a 
sub-solar mass compact object incompatible with a \ac{ns}. The biases in the chirp mass and mass ratio reflect 
on the inference of the reduced tidal parameters which, in the considered experiment, is significantly underestimated.
This would have obvious consequences for inference of the \ac{ns} matter. 
However, we also find that the parameters of eccentric binaries with $e_0\lesssim0.01$ are robustly determined by 
a quasicircular and spinning waveform at \ac{snr} ${\lesssim}32$. Under these conditions, eccentric nonspinning 
\ac{bns} signals appear well reproduced by nonprecessing quasicircular templates, but higher \acp{snr} are expected 
to introduce again significant biases.

Our case study focuses on an unequal-mass \ac{bns} with intrinsic parameters compatible with GW170817, including 
the considered eccentricities \cite{Lenon:2020oza}.
We performed full Bayesian \acp{pe} of mock signals at \ac{snr} $12.7, \ 15.9, \ 31.7$ using the 
state-of-art \ac{eob} model \TEOB{-\dali} with either eccentricity or spin precessional effects.
We find that quasicircular waveform determines fractional \ac{snr} losses from a few to ${\sim}25$\% for 
eccentricities from $e=0.01$ to $0.1$. The \ac{snr} losses have an approximate linear dependence with 
eccentricity in our data, see Eq.~\eqref{eq:snr-loss-fit}.
The chirp mass is significantly biased toward larger values as the eccentricity increases. This bias can 
be understood in terms of the eccentric chirp mass parameter $\Mecc$ introduced in \cite{Favata:2021vhw} 
(our Eq.~\eqref{eq:mchirp-ecc}). Indeed, for low eccentricities we recover values of $\M$ that are consistent 
with $\Mecc$ within their 90\% \ac{ci}. The eccentric chirp mass argument however does not provide a 
quantitative prediction for the bias for higher eccentricities $e_0\gtrsim0.1$. On the one hand, these 
eccentricities are out of the regime of validity of Eq.~\eqref{eq:mchirp-ecc}; on the other hand there is 
a strong degeneracy with the mass ratio.
The mass ratio
is degenerate with the eccentric chirp mass and can reach
unphysically large values for \ac{bns} to reproduce the eccentricity
effect. This effect can hamper real-signal analyses and lead to
an incorrect interpretation of the source. Large mass ratios
would favor a \ac{bh} interpretation for the primary compact object
and a unknown companion, even if the reduced tidal parameter can
be measured.
We cannot confirm the simple quadratic scaling 
with eccentricity of \citet{Cho:2022cdy} for the chirp mass bias because our bias is strongly degenerated 
with the mass ratio.
The effective spin parameter is always inferred as positive and up $\chi_{\rm eff}\sim 0.8$ for $e_0\lesssim0.1$. 
This suggests that repulsive spin-orbits interactions are effectively counteracting the attractive effect of 
the larger chirp mass and reduced tidal parameter for the fitting of the eccentric data.
The inclusion of spin precession in the waveform is also insufficient to capture eccentricity effects. 
Therefore, we conclude that the spin precession cannot mimic eccentricity for long signals as 
in our analyses since the effects have two different timescales.
Furthermore, this case study suggests that the eccentricity of GW170817 is
$e_0 < 0.05$, confirming \cite{Lenon:2020oza}, otherwise unphysically large values 
of the mass ratio may have been observed.
In the \ac{pe} of GW170817, two different priors for the spins have been considered \cite{TheLIGOScientific:2017qsa}, while in all our analyses,
we adopt the high-spin priors. When employing low-spin priors, we expect the posterior distributions of effective spin parameters to rail
against the priors and a larger \ac{snr} loss compared to the high-spin priors case. 

We conclude that high-precision observations with both upcoming advanced \cite{LIGOScientific:2014pky,VIRGO:2014yos,KAGRA:2013rdx} and 
next generation~\cite{Hild:2010id,Hild:2011np,Punturo:2010zz,Branchesi:2023mws,Abac:2025saz,Reitze:2019iox}
detectors are likely to require standardized and accurate eccentric \ac{bns} waveforms.
While the latter are available in \TEOB{-\dali}, the computational cost of waveform generation \emph{and} 
inference should be improved in the near future. In the eccentric case, \TEOB{-\dali} cannot yet take advantage 
of acceleration methods like the post-adiabatic \cite{Nagar:2018gnk} and EOB-SPA \cite{Gamba:2020ljo}, although 
this effect can be incorporated in reduced-order/machine learning models~\cite{Tissino:2022thn}. On the 
inference side, speed up techniques such as relative binning \cite{Dai:2018dca, Zackay:2018qdy} cannot 
be employed yet, because a frequency domain description of the model is needed. We could instead consider 
to use reduced order quadrature \cite{Canizares:2014fya,Smith:2016qas,Morisaki:2020oqk}
or RIFT \cite{Pankow:2015cra,Lange:2018pyp,Wysocki:2019grj,Wofford:2022ykb}.
  
Future work will be focused on a more systematic exploration of different \ac{bns} configurations and 
waveform effects. An important aspect neglected here is the inclusion of higher modes which may enhance 
the performance of the spin-precessing waveform in capturing the eccentric feature of the signal. Higher 
modes are available in \TEOB{-\dali} but increase the cost of the inference. Similarly, work is ongoing in 
exploring inference with eccentric and spinning waveform. Again, there is no conceptual difficulty in 
performing these analyses with \TEOB{-\dali} but their increased computational cost.

\begin{acknowledgments}
The authors thank Gregorio Carullo, Axel Kuch, Jake Lange and Nicolo' Venuti for early contributions and discussions on this project.
G.H. and S.B. acknowledge support by the EU Horizon under ERC Consolidator Grant, No. InspiReM-101043372.
S.B. and M.B. acknowledge support by the EU H2020 under ERC Starting Grant, No.~BinGraSp-714626.  
M.B. and R.G. acknowledge support by the Deutsche Forschungsgemeinschaft (DFG) under Grant No. 406116891 within the Research Training Group RTG 2522/1.
R.G. acknowledges support from NSF Grant PHY-2020275 [Network for Neutrinos, Nuclear Astrophysics, and Symmetries (N3AS)].

This research has made use of data obtained from the Gravitational Wave Open 
Science Center \cite{lvk_data}, a service of LIGO
Laboratory, the LIGO Scientific Collaboration, the
Virgo Collaboration, and KAGRA. This material is based upon work supported
by NSF's LIGO Laboratory, which is a major facility fully funded by
the National Science Foundation. LIGO Laboratory
and Advanced LIGO are funded by the United States
National Science Foundation (NSF) as well as the Science and 
Technology Facilities Council (STFC) of the
United Kingdom, the Max-Planck-Society (MPS), and
the State of Niedersachsen/Germany for support of the
construction of Advanced LIGO and construction and
operation of the GEO600 detector. Additional support
or Advanced LIGO was provided by the Australian Research Council. 
Virgo is funded through the European
Gravitational Observatory (EGO), by the French Centre
National de Recherche Scientifique (CNRS), the Italian
Istituto Nazionale di Fisica Nucleare (INFN), and the
Dutch Nikhef, with contributions by institutions from
Belgium, Germany, Greece, Hungary, Ireland, Japan,
Monaco, Poland, Portugal, Spain. KAGRA is supported
by Ministry of Education, Culture, Sports, Science, and
Technology (MEXT), Japan Society for the Promotion
of Science (JSPS) in Japan; National Research Foundation (NRF) 
and Ministry of Science and ICT (MSIT)
in Korea; Academia Sinica (AS) and National Science
and Technology Council (NSTC) in Taiwan.

Simulations were performed on SuperMUC-NG at the Leibniz-Rechenzentrum
(LRZ) Munich and on the national HPE Apollo Hawk at the High
Performance Computing Center Stuttgart (HLRS). 
The authors acknowledge the Gauss Centre for Supercomputing
e.V. (\url{www.gauss-centre.eu}) for funding this project by providing
computing time on the GCS Supercomputer SuperMUC-NG at LRZ
(allocations {\tt pn36go}, {\tt pn36jo} and {\tt pn68wi}). The authors
acknowledge HLRS for funding this project by providing access to the
supercomputer HPE Apollo Hawk under the grant number INTRHYG UE/44215
and MAGNETIST/44288. 
Computations were also performed on the ARA cluster at Friedrich
Schiller University Jena and on the {\tt Tullio} INFN cluster at INFN
Turin. The ARA cluster is funded in part by DFG Grants INST 275/334-1
FUGG and INST 275/363-1 FUGG, and ERC Starting Grant, Grant Agreement
No. BinGraSp-714626. 

\section{Data availability}
The data that support the findings of this article are not publicly available. 
The data are available from the authors upon reasonable request. 

\TEOB{} is publicly available at
  
  {\footnotesize \url{https://bitbucket.org/teobresums}}.
  
{\bajes} is publicly available at

  {\footnotesize \url{https://github.com/matteobreschi/bajes}}.
 
\end{acknowledgments}

\end{document}